\documentclass[prc,floatfix,groupedaddress,nofootinbib,showpacs,preprintnumbers,
amsmath,amssymb,amsfonts,superscriptaddress,widetable] {revtex4-1}
\usepackage{graphicx}
\usepackage{dcolumn}
\usepackage{mathrsfs}
\usepackage{bm}
\usepackage{float}
\usepackage{soul}
\usepackage[usenames]{color}

\newcommand{\rhozero}{\rho_{\raisebox{-2.0pt}{\tiny\!0}}}
\newcommand{\epszero}{\varepsilon_{\raisebox{-2.0pt}{\tiny\!0}}}
\newcommand{\nzero}{n_{\raisebox{-1pt}{\tiny0}}}
\newcommand{\pFermi}{p_{\raisebox{-1pt}{\tiny{\rm F}}}}

\begin{document}

\begin{titlepage}
\begin{center}

\includegraphics[height=4cm]{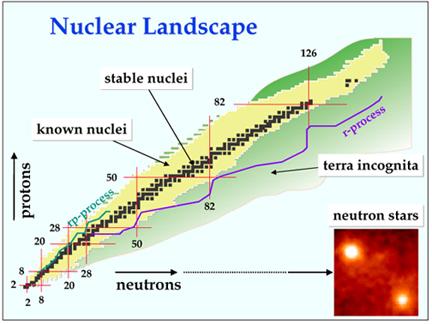}
\includegraphics[height=4cm]{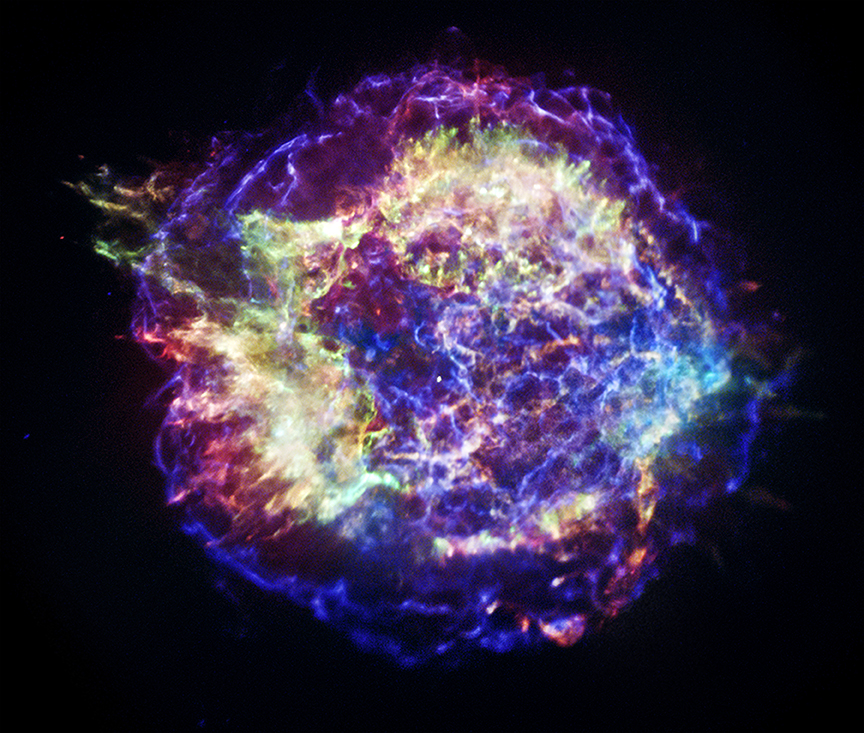}~\\[1cm]

{\huge \bfseries Relativistic density functional theory for \\ 
finite nuclei and neutron stars\footnote{Contributing chapter to the book ``Relativistic Density 
Functional for Nuclear Structure"; World Scientific Publishing Company (Singapore); Editor 
Prof. Jie Meng.} \\[0.5cm] }
\textsc{\LARGE J. Piekarewicz}\\[0.3cm]
\textsc{\Large Department of Physics}\\[0.2cm]
\textsc{\Large Florida State University}\\[0.2cm] 
\textsc{\Large Tallahassee, FL 32306}\\[0.5cm]
{\large \today}\\[0.5cm]
\end{center}

\vspace{1cm}
\begin{abstract}
{\large
{\bf Abstract:} The main goal of the present contribution is a pedagogical introduction to the fascinating 
world of neutron stars by relying on relativistic density functional theory. Density 
functional theory provides a powerful---and perhaps unique---framework for the calculation of both 
the properties of finite nuclei and neutron stars. Given the enormous densities that may be reached
in the core of neutron stars, it is essential that such theoretical framework incorporates from the 
outset the basic principles of Lorentz covariance and special relativity. After a brief historical
perspective, we present the necessary details required to compute the equation of state of dense,
neutron-rich matter. As the equation of state is all that is needed to compute the structure of neutron
stars, we discuss how nuclear physics---particularly certain kind of laboratory experiments---can 
provide significant constrains on the behavior of neutron-rich matter.}
\end{abstract}
\maketitle

\end{titlepage}


\section{Introduction}
\label{Introduction}

\begin{figure}[h]
\begin{center}
  \includegraphics[width=0.4\columnwidth]{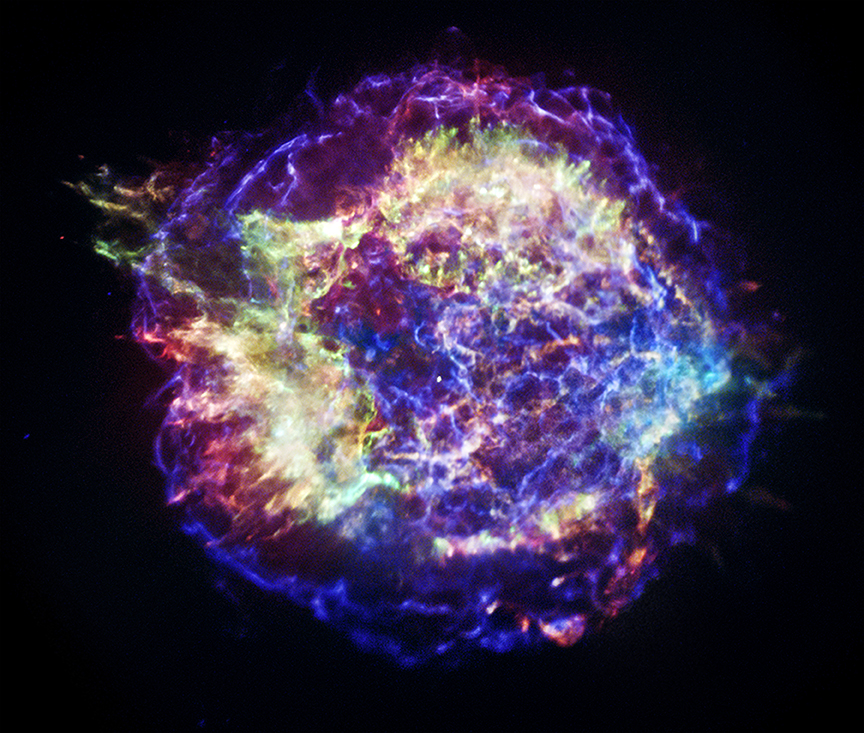}
  \vspace{-0.2cm}
  \caption{Cassiopeia A is the remnant of a supernovae explosion that
  was witnessed in the northern sky about 300 years ago. The supernova
  remanent is located about 10,000 light years away from Earth and the
  small ``dot'' near the center of the image represents the dense neutron 
  star. The image was created using NASA's Chandra x-ray observatory,
  an observatory named in honor of Subramanyan Chandrasekhar.}
 \label{FigIntro1}
\end{center}
\end{figure}

The birth of a star is marked by the conversion of hydrogen into
helium nuclei ($\alpha$ particles) in their hot dense cores.  This
thermonuclear reaction is the main source of energy generation during
the main stage of stellar evolution and provides the pressure support
against gravitational collapse.  Once the hydrogen in the stellar core
is exhausted, thermonuclear fusion stops and the star contracts. As a
result of the gravitational contraction, the temperature in the
stellar core increases to about 100 million K allowing the heavier
helium ashes to overcome their electrostatic repulsion and fuse into
heavier elements. However, the absence of stable nuclei containing
either five or eight nucleons hinders the production of heavy
elements. Remarkably, the conditions of density and temperature in the
stellar interior are such that a minute equilibrium concentration of
${}^{8}$Be develops; the concentration of ${}^{8}$Be relative to that
of ${}^{4}$He is about 1 parts per billion!  Yet this minute
concentration is sufficient for another
$\alpha$-particle to be captured leading to the formation of a
${}^{12}$C nucleus. The physics of the ``triple-alpha'' reaction,
including the prediction of the resonant Hoyle state, is one of the
most fascinating chapters in the story of stellar
nucleosynthesis\,\cite{Clayton:1983,Iliadis:2007}. Although in stars
as our Sun the formation of heavier elements is hindered by the
degeneracy pressure of the electrons, the conditions in the core of
more massive stars are conducive to the formation of heavier elements,
such as ${}^{16}$O, ${}^{24}$Mg, ${}^{28}$Si, ${}^{32}$S.  However,
abruptly and unavoidably, the fusion of light nuclei into ever
increasing heavier elements terminates with the synthesis of the
iron-group elements (Fe, Co, and Ni) that are characterized by having
the largest binding energy per nucleon. That is, once iron-group
elements are produced in the stellar interior, it is no longer feasible to
generate energy by thermonuclear fusion.  Indeed, if the iron core
exceeds the ``Chandrasekhar limit" of about 1.4 solar masses, neither
thermonuclear fusion nor electron degeneracy pressure can prevent the
rapid collapse of the stellar core. The collapse of the core, with the
ensuing shock wave that disseminates the chemical elements crafted
during the lifetime of the star, produces one of the most remarkable
events in the Universe: a \emph{Supernova Explosion}
(see Fig.\ref{FigIntro1}). Besides creating an ejecta that contains some 
of the essential elements necessary for life, core-collapse supernovae 
leave behind exotic compact remnants in the form of either black holes 
or \emph{neutron stars}. Neutron stars are the central theme of the 
present contribution.

\vspace{-0.1cm}

Indirectly and inadvertently, Subramanyan Chandrasekhar (``Chandra'')
may have been the discoverer of neutron stars. In a pioneering paper 
published in 1931, Chandra re-examined the role of electron degeneracy 
pressure in supporting a white-dwarf star against gravitational collapse, 
a fact that was already well known at the time. Chandra realized, however, 
that as the electrons become relativistic, the pressure support weakens 
and a white-dwarf star with a mass in excess of about 1.4 solar masses 
(the so-called ``Chandrasekhar mass limit'') will collapse under its own
weight\,\cite{Chandrasekhar:1931}. Chandra summarized eloquently this
critical finding: \emph{For a star of small mass the white-dwarf stage
is an initial step towards complete extinction. A star of large mass
cannot pass into the white-dwarf stage and one is left speculating on
other possibilities.} One may ask why did Chandra never speculated
that neutron stars may be among the ``other possibilities". As luck
will have it, the neutron was not yet discovered in 1931; it would
take Chadwick another year to announce the 
discovery\,\cite{Chadwick:1932}. However, soon after Chadwick's
announcement, the term \emph{neutron star} appears in writing for the
first time in the 1933 proceedings of the the American Physical
Society by Baade and Zwicky who wrote: \emph{With all reserve we
advance the view that supernovae represent the transition from
ordinary stars into ``neutron stars'', which in their final stages
consist of extremely closed packed neutrons}\,\cite{Baade:1934}.

It appears, however, that speculations on the possible existence of both 
the neutron and neutron stars may have an earlier origin. In the case of 
the neutron, the story starts with the ``boys of Via Panisperna'' who included 
such luminaries as Enrico Fermi, Franco Rasetti, and Ettore Majorana, among 
others. During a brief stay at Caltech (in the 1928-29 period) working 
in Millikan's laboratory, Rasetti measured the ground-state angular 
momentum of ${}^{14}$N to be $J\!=\!1$. At the time, the widespread 
believe was that the nucleus of ${}^{14}$N must contain 14 protons (to 
account for its mass) and 7 electrons (to account for its charge). 
However, Majorana was the first one to realize that 21 spin-1/2 
fermions can not account for the measured spin of ${}^{14}$N, 
thereby postulating the existence of an electrically neutral spin-1/2 particle 
having the same mass as the proton\,\cite{Magueijo:2009}. Moreover, 
the requirement for such a particle fitted correctly Majorana's interpretation 
of some experiments carried out in 1932 by Ir\`ene Joliot-Curie and 
Fr\'ed\'eric Joliot. Apparently Fermi pleaded with Majorana to write an 
article on the neutron, but Majorana did not find it worthy.

\begin{figure}[h]
\begin{center}
 \includegraphics[height=7cm]{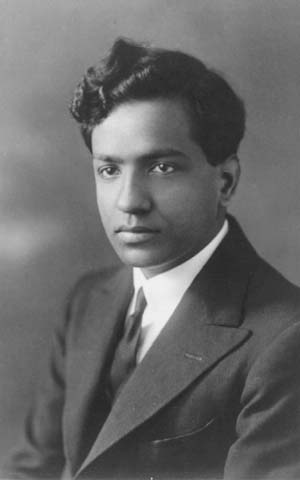}
 \includegraphics[height=7cm]{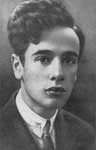} 
 \vspace{-0.2cm}
   \caption{Subrahmanyan Chandrasekhar (left) and Lev Landau (right) who 
   wrote seminal papers on the theory of stars at the ``ripe'' age of 19 and 23,
   respectively.}
 \label{FigIntro2}
\end{center} 
\end{figure}

Insofar as Landau's role on the history of neutron stars is concerned, the first 
recollection appears on a paper entitled ``On the theory of stars'' that Landau 
submitted for publication in early 1932 at the age of 23\,\cite{Landau:1932}. In 
that paper Landau calculates---independently of Chandrasekhar---the maximum 
mass of a white dwarf star. Moreover, unlike Chandrasekhar, Landau did 
speculate on ``other possibilities'', namely, the existence of dense stars that 
look like giant atomic nuclei. For an in depth and fascinating tale on Landau's 
role on the possible existence of neutron stars see Ref.\,\cite{Yakovlev:2012rd}. 
Note that in Fig.\,\ref{FigIntro2} we display pictures of both Chandra and Landau 
as very young men. 

Perhaps the last great theoretical landmark of that time involves 
the 1939 work by Oppenheimer and Volkoff on the structure of neutron
stars\,\cite{Opp39_PR55}. By then, Einstein's general theory of
relativity was firmly established as was Tolman's framework to compute
solutions appropriate to spherical systems in hydrostatic
equilibrium\,\cite{Tol39_PR55}. In what it is now referred to as the
Tolman-Volkoff-Oppenheimer (TOV) equations---effectively the
generalization of Newtonian gravity to the domain of general
relativity---Oppenheimer and Volkoff concluded that a neutron star
supported exclusively by the pressure from its degenerate neutrons
will collapse into (what we now know as) a black hole for masses in
excess of about 0.7 solar masses. This critical finding, together with 
our present knowledge of neutron-star masses, has made nuclear 
physics and astrophysics intimately intertwined.
\begin{figure}[h]
\begin{center}
 \includegraphics[height=9cm]{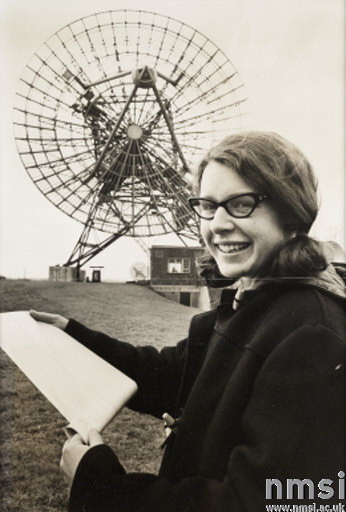}
 \vspace{-0.2cm}
   \caption{Jocelyn Bell as a young graduate student with her radio
    telescope designed to search for quasars. Instead, and according 
    to Dr. Iosif Shklovsky, she made the  greatest astronomical 
    discovery of the twentieth century.}
 \label{FigIntro3}
\end{center} 
\end{figure}

Although firmly established theoretically, it would take almost three
decades for the discovery of neutron stars. This momentous discovery
started with a young graduate student by the name of Jocelyn 
Bell (see Fig.\,\ref{FigIntro3})---now \emph{Dame} Jocelyn Bell 
Burnell---who detected a ``bit of scruff'' in the data arriving into her 
radio telescope, a telescope originally designed to study distant 
quasars. The arriving signal was ``pulsing'' with such an enormous
regularity, indeed once every 1.337\,302\,088\,331 seconds that both
Bell and her research advisor, Anthony Hewish, were so mesmerized 
by the observation that they were convinced that the signal was a 
beacon from an extraterrestrial civilization. Initially dubbed as 
``Little Green Man 1'' the source, now known as radio pulsar 
``PSR B1919+21'', was shortly identified as a rapidly rotating neutron
star\,\cite{Hewish:1968}. Although Hewish was recognized with the
Physics Nobel Prize in 1974 for ``his decisive role in the discovery of
pulsars'', Jocelyn Bell did not share the award. The exclusion of
Jocelyn Bell as co-recipient of the Nobel Prize was both controversial
and roundly condemned by the astrophysics community. Still, Bell has
always displayed enormous grace and humility in the face of this
controversy. Indeed, she has stated: \emph{I believe it would demean
Nobel Prizes if they were awarded to research students, except in very
exceptional cases, and I do not believe this is one of them.}  It
appears that Dr. Iosif Shklovsky, as well as many others, did not
share her views. Dr. Shklovsky---recipient of the 1972 Bruce Medal for
outstanding lifetime contributions to astronomy---told Jocelyn Bell:
\emph{Miss Bell, you have made the greatest astronomical discovery 
of the twentieth century.}
\begin{table}[!htbp]
\caption{Approximate characteristics of a ``canonical'' neutron star, such as the 
960 year old Crab pulsar.}
{\begin{tabular}{@{}ll@{}} \toprule
  Name: PSR B0531+21 & Constellation: Taurus\\
  Distance: 2.2 kpc  & Age: 960 years\\
  Mass: $1.4\,M_{\odot}$ & Radius: 10\,km \\ 
  Density: $10^{15}$g/cm${}^{3}$ & Pressure: $10^{29}$\,atm \\  
  Surface Temperature: $10^{6}$\,K & Escape velocity: 0.6\,c\\
  Period: 33\,ms & Magnetic Field: $10^{12}$ G \\ \botrule
\end{tabular}}  
\label{TableIntro1}
\end{table}    

The finding by Oppenheimer and Volkoff of a maximum neutron star mass
of about 0.7 solar masses\,\cite{Opp39_PR55} will eventually place
nuclear physics at the forefront of neutron-star structure. This is because 
the masses of various neutron stars have been determined very accurately
and they exceed, often by a significant margin, the 0.7\,$M_{\odot}$ limit 
(see Ref.\,\cite{Lattimer:2012nd} and references contained therein). Indeed, 
the present ``record'' stands at 
2\,$M_{\odot}$\,\cite{Demorest:2010bx,Antoniadis:2013pzd}. Given that
the Oppenheimer-Volkoff limit was obtained by assuming that the
support against gravitational collapse is provided by a degenerate gas
of neutrons, the large deficit must then be supplied by nuclear
interactions. Thus, neutron stars are enormously valuable in
constraining the largely undetermined equation of state of
neutron-rich matter at high densities. However, the reach of neutron
stars is not limited to the nuclear physics domain. Indeed, neutron
stars are unique laboratories for the study of matter under extreme
conditions of density and isospin asymmetry. In particular, their
extreme compactness has been used to test the basic tenets of  
general relativity\,\cite{Hulse:1974eb}. Moreover, by spanning many 
orders of magnitude in density, neutron stars display exotic phases 
that cannot be realized under normal laboratory conditions. Yet, 
some of these phases have direct counterparts in both atomic physics 
and condensed-matter physics.  Finally, their very dense stellar cores 
may harbor novel states of matter, such as color superconductors, that 
are a direct prediction of Quantum Chromodynamics\,\cite{Alford:1998mk}. 
To appreciate some of the unique properties of neutron stars, such as 
their density, pressure, spin period, and magnetic fields, we have 
listed some of these properties in Table\,\ref{TableIntro1} for the case 
of the well-known crab pulsar.

We have organized this chapter on neutron stars as follows. After
this Introduction, we present the formalism that will be used to 
compute some structural properties of neutron stars, such as their 
mass-vs-radius relation. Once the formalism is in place, we embark 
on a journey of a neutron star that involves a detailed discussion of
the outer crust, the inner crust, and the outer core. For all these three
cases we present results that highlight those stellar observables that are 
particularly sensitive to the choice of the nuclear density functional. 
We finish with a summary of our results and an outlook for the future.

\section{Formalism}
\label{formalism} 

The ultimate goal of a properly constructed nuclear energy density
functional is to provide a unified description of a wide variety of
physical phenomena ranging from the properties of finite nuclei to 
the structure and dynamics of neutron stars. This ambitious goal
involves physical systems that differ in mass and size by about 
55 and 18 orders of magnitude, respectively. Moreover, given that 
extrapolations into regions that are inaccessible in laboratory 
experiments are unavoidable, the predictions of such microscopic 
theory should always be accompanied by well-quantified theoretical 
uncertainties\,\cite{Kortelainen:2010hv,PhysRevA.83.040001}.  

Historically, relativistic models of nuclear structure were limited to 
renormalizable field theories\,\cite{Walecka:1974qa,Serot:1979dc}.
The appeal of renormalizability was evident: with only a handful of 
model parameters calibrated to well-known physical observables
one could then extrapolate to unknown physical regions without 
the need for introducing additional parameters. However, the modern 
viewpoint suggests that any relativistic model, although often inspired 
in Quantum Chromodynamics (QCD), should be treated as an 
\emph{effective field theory} (EFT) where the demand for renormalizability 
is no longer required. An effective field theory is designed to describe 
low-energy physics without any attempt to account for its detailed 
short-distance behavior\,\cite{Furnstahl:2000in}. Although in principle 
the empirical parameters of the EFT may be calculable from QCD, in 
practice this becomes enormously challenging in the non-perturbative 
regime of relevance to nuclear systems. Hence, the parameters of the 
model are directly calibrated (i.e., fitted) to physical observables. By 
doing so, the short-distance structure of the theory as well as many other 
complicated many-body effects get implicitly encoded in the parameters 
of the model. In this regard, density functional theory (DFT), a powerful
and highly successful theoretical framework pioneered by Kohn and 
collaborators\,\cite{Hohenberg:1964zz,Kohn:1965,Kohn:1999},  provides 
a unified approach for the construction of an EFT that may be used to 
compute phenomena ranging over many distance scales.

\subsection{Relativistic Density Functional Theory}
\label{RDFT}

In the framework of the relativistic density functional theory, the effective
degrees of freedom include nucleons (protons and neutrons), three 
``mesons'', and the photon. The interactions among the particles can be 
described as a generalization of the original Lagrangian density of Serot
and Walecka\,\cite{Walecka:1974qa,Serot:1984ey, Mueller:1996pm,
Serot:1997xg, Horowitz:2000xj}. That is,
\begin{eqnarray}
 {\mathscr L}_{\rm int} &=&
\bar\psi \left[g_{\rm s}\phi   \!-\! 
         \left(g_{\rm v}V_\mu  \!+\!
    \frac{g_{\rho}}{2}{\mbox{\boldmath $\tau$}}\cdot{\bf b}_{\mu} 
                               \!+\!    
    \frac{e}{2}(1\!+\!\tau_{3})A_{\mu}\right)\gamma^{\mu}
         \right]\psi \nonumber \\
                   &-& 
    \frac{\kappa}{3!} (g_{\rm s}\phi)^3 \!-\!
    \frac{\lambda}{4!}(g_{\rm s}\phi)^4 \!+\!
    \frac{\zeta}{4!}   g_{\rm v}^4(V_{\mu}V^\mu)^2 \nonumber \\ &+&
   \Lambda_{\rm v}\Big(g_{\rho}^{2}\,{\bf b}_{\mu}\cdot{\bf b}^{\mu}\Big)
                           \Big(g_{\rm v}^{2}V_{\nu}V^{\nu}\Big)\;,
 \label{LDensity}
\end{eqnarray}
where $\psi$ is the isodoublet nucleon field, $A_{\mu}$ is the 
photon field, and $\phi$, $V_{\mu}$, and ${\bf b}_{\mu}$ represent 
the isoscalar-scalar $\sigma$-, isoscalar-vector $\omega$-, and
isovector-vector $\rho$-meson field, respectively. We note that
the pion is not explicitly included in the Lagrangian density as it 
does not contribute to the nuclear dynamics at the mean field level. 
The Lagrangian density incorporates the conventional Yukawa terms 
between the nucleon and the various mesons and the photon. However, 
in order to improve the quality of the model it is critical to supplement
the dynamics with nonlinear interaction terms between the various mesons. 
In the spirit of an effective field theory, one should 
incorporate all possible meson interactions that are allowed by symmetry
considerations to a given order in a power-counting scheme. Moreover,
once the \emph{dimensionful} meson fields have been properly scaled 
using strong-interaction mass scales, the remaining dimensionless
coefficients of the effective Lagrangian should all be ``natural'', namely,
neither too small nor too large\,\cite{Furnstahl:1996wv,Furnstahl:1996zm,
Rusnak:1997dj,Furnstahl:1997hq,Kortelainen:2010dt}. However,
given the limited experimental database of nuclear observables, 
certain empirical coefficients---or linear combinations of 
them---may remain poorly determined even after the optimization
procedure. This results in ``unnatural''  coefficients that deviate 
significantly from unity. Therefore, in an effort to avoid this problem, 
only those nonlinear meson interactions with a clear physical 
interpretation are retained. For instance, for the Lagrangian density 
depicted in Eq.\,(\ref{LDensity}), we have only kept the four 
non-linear meson interactions that are denoted by the coefficients: 
$\kappa$, $\lambda$, $\zeta$, and $\Lambda_{\rm v}$. Two of the 
isoscalar parameters, $\kappa$ and $\lambda$, were introduced by 
Boguta and Bodmer\,\cite{Boguta:1977xi} to soften the equation of
state of symmetric nuclear matter, primarily the incompressibility
coefficient\,\cite{Walecka:1974qa,Serot:1984ey}, in an effort to
make the theory consistent with measurements of giant monopole 
resonances in finite nuclei. In turn, $\zeta$ may be used to efficiently 
tune the maximum neutron star mass without sacrificing the agreement 
with other well reproduced observables\,\cite{Mueller:1996pm}. Finally, 
$\Lambda_{\rm v}$ is highly sensitive to the density dependence of 
symmetry energy---and in particular to its slope at saturation 
density---which has important implications in the structure and
dynamics of neutron stars\,\cite{Horowitz:2000xj,Horowitz:2001ya,
Carriere:2002bx,Horowitz:2004yf}. 

With the Lagrangian density given in Eq.\,(\ref{LDensity}), one can 
derive the equations of motion for each of the constituent particles 
in the mean-field approximation\,\cite{Todd:2003xs}. In particular,
the nucleons satisfy a Dirac equation in the presence of mean-field 
potentials of Lorentz scalar and vector character. In turn, the various 
meson fields satisfy both nonlinear and inhomogeneous Klein-Gordon 
equations with the various nuclear densities acting as source terms. 
Given that the nuclear densities act as sources for the meson fields 
and, in turn, the meson fields determine the mean-field potentials for 
the nucleons, the set of equations must be solved self-consistently. 
Once solved, these equations determine the  ground-state properties 
of the nucleus of interest---such as its total binding energy, the
single-nucleon energies and Dirac orbitals, the distribution of meson 
fields, and the various density profiles.

\subsection{Nuclear Matter Equation of State}
\label{EOS}

The solution of the mean-field equations is simplified considerably in 
the case of infinite nuclear matter, which is assumed to be spatially 
uniform. Although the solution may be found at finite temperature, 
our main goal is to solve the mean-field equations at zero temperature
because of their relevance to the structure and dynamics of neutron
stars. Indeed, if the validity of Einstein's theory of General Relativity
is assumed, then the equation of state of asymmetric nuclear matter
represents the sole ingredient required to compute the properties of 
neutron stars (see next section). 

In the simplified case of infinite nuclear matter, the meson fields are 
uniform (i.e., constant throughout space) and the nucleon orbitals are 
plane-wave Dirac spinors with medium-modified effective masses and 
energies that must be determined self-consistently. By constructing 
the energy-momentum tensor in the mean-field 
approximation\,\cite{Serot:1984ey}, one obtains the equation of state 
of asymmetric nuclear matter, namely, the energy density and pressure 
of the system as a function of both the conserved baryon density 
$\rho\!=\rho_{n}\!+\!\rho_{p}$ and the neutron-proton asymmetry
$\alpha\!\equiv\!(\rho_{n}\!-\!\rho_{p})/(\rho_{n}\!+\!\rho_{p})$. A
particularly insightful view of the EOS is obtained by expanding the 
energy per nucleon in powers of the neutron-proton asymmetry. That 
is, 
\begin{equation}
  \frac{E}{A}(\rho,\alpha) -\!M \equiv {\cal E}(\rho,\alpha)
                          = {\cal E}_{\rm SNM}(\rho)
                          + \alpha^{2}{\cal S}(\rho)  
                          + {\cal O}(\alpha^{4}) \,.
 \label{EOS}
\end {equation}
Here ${\cal E}_{\rm SNM}(\rho)\!=\!{\cal E}(\rho,\alpha\!\equiv\!0)$
is the energy per nucleon of symmetric nuclear matter (SNM) and 
the symmetry energy ${\cal S}(\rho)$ represents the first-order 
correction to the symmetric limit. Note that no odd powers of $\alpha$ 
appear as the nuclear force is assumed to be isospin symmetric and 
long-range electromagnetic effects have been ``turned off''. Also note 
that to a very good approximation the symmetry energy represents the 
energy cost required to convert symmetric nuclear matter into pure 
neutron matter (PNM). That is,
\begin{equation}
 {\cal S}(\rho)\!\approx\!{\cal E}(\rho,\alpha\!=\!1) \!-\! 
 {\cal E}(\rho,\alpha\!=\!0) \;.
 \label{SymmE}
\end {equation}
Such a separation is useful because symmetric nuclear matter is
sensitive to the isoscalar sector of the density functional which is 
well constrained by the properties of stable nuclei. In contrast, the 
symmetry energy probes the isovector sector of the density functional 
which at present is poorly constrained. However, this problem will be
mitigated with the commissioning of radioactive beam facilities 
throughout the world.

Besides the separation of the EOS into symmetric and asymmetric 
components, it is also useful to characterize the behavior of the 
equation of state in terms of a few bulk parameters. To do so one 
performs a Taylor series expansion around nuclear matter saturation 
density $\rhozero$. That is\,\cite{Piekarewicz:2008nh},
\begin{subequations}
\begin{align}
 & {\cal E}_{\rm SNM}(\rho) = \epszero + \frac{1}{2}K_{0}x^{2}+\ldots ,\label{EandSa}\\
 & {\cal S}(\rho) = J + Lx + \frac{1}{2}K_{\rm sym}x^{2}+\ldots ,\label{EandSb}
\end{align} 
\label{EandS}
\end{subequations}
\!\!\!where $x\!=\!(\rho-\rhozero)\!/3\rhozero$ is a dimensionless parameter 
that quantifies the deviations of the density from its value at saturation. 
Here $\epszero$ and $K_{0}$ represent the energy per nucleon and the 
incompressibility coefficient of SNM; $J$ and $K_{\rm sym}$ are the 
corresponding quantities for the symmetry energy. However, unlike 
symmetric nuclear matter whose pressure vanishes at $\rhozero$, the 
slope of the symmetry energy $L$ does not vanish at saturation density. 
Indeed, assuming the validity of Eq.\,(\ref{SymmE}), $L$ is directly 
proportional to the pressure of PNM ($P_{0}$) at saturation density, 
namely,
\begin{equation}
   P_{0} \approx \frac{1}{3}\rhozero L \;.
 \label{PvsL}
\end{equation}
In computing various neutron-star observables in the next few sections, 
we will rely on several nuclear density functionals that while successful 
in reproducing a myriad of laboratory observables, predict significant
differences in the properties of neutron stars.

\subsection{Tolman-Oppenheimer-Volkoff Equations}
\label{TOV}

With masses comparable to that of our Sun but with radii that are 
almost five orders of magnitude smaller (i.e., of the order of 10 km) 
neutron stars are highly compact objects that must be described 
using Einstein's theory of General Relativity. The generalization 
of Newtonian gravity to the realm of general relativity is expressed
in the \emph{Tolman-Oppenheimer-Volkoff} (TOV) equations, which 
are usually presented as a coupled set of first-order differential 
equations of the following form:
 \begin{eqnarray}
   && \frac{dP}{dr}\!=\!-\!G\,\frac{{\cal E}(r)M(r)}{r^{2}}
         \left[1\!+\!\frac{P(r)}{{\cal E}(r)}\right]
         \left[1\!+\!\frac{4\pi r^{3}P(r)}{M(r)}\right]
         \left[1\!-\!\frac{2GM(r)}{r}\right]^{-1}\hspace{-12pt},
         \label{TOVa}  \\
   && \frac{dM}{dr}=4\pi r^{2}{\cal E}(r)\;,
         \label{TOVb}
 \label{TOV}
\end{eqnarray}
where $G$ is Newton's gravitational constant and $P(r)$, ${\cal E}(r)$, and $M(r)$ 
represent the pressure, energy density, and enclosed-mass profiles of the star, 
respectively. Note that the  three terms enclosed in square brackets in Eq.\,(\ref{TOVa}) 
are of general-relativistic origin. As already alluded earlier and particularly interesting, 
the only input that neutron stars are sensitive to is the equation of state of
neutron-rich matter. This fact alone creates a unique synergy between nuclear
physics and astrophysics. In essence, by specifying the central pressure and 
enclosed mass, i.e., $P_{c}\!=\!P(r\!=\!0)$ and $M(r\!=\!0)\!=\!0$---together with 
a suitable EOS---the TOV equations may be solved using a standard numerical 
algorithm, such as the Runge-Kutta method. 

\section{Anatomy of a Neutron Star}
\label{Anatomy}
\begin{figure}[h]
\begin{center}
  \includegraphics[height=2.25in]{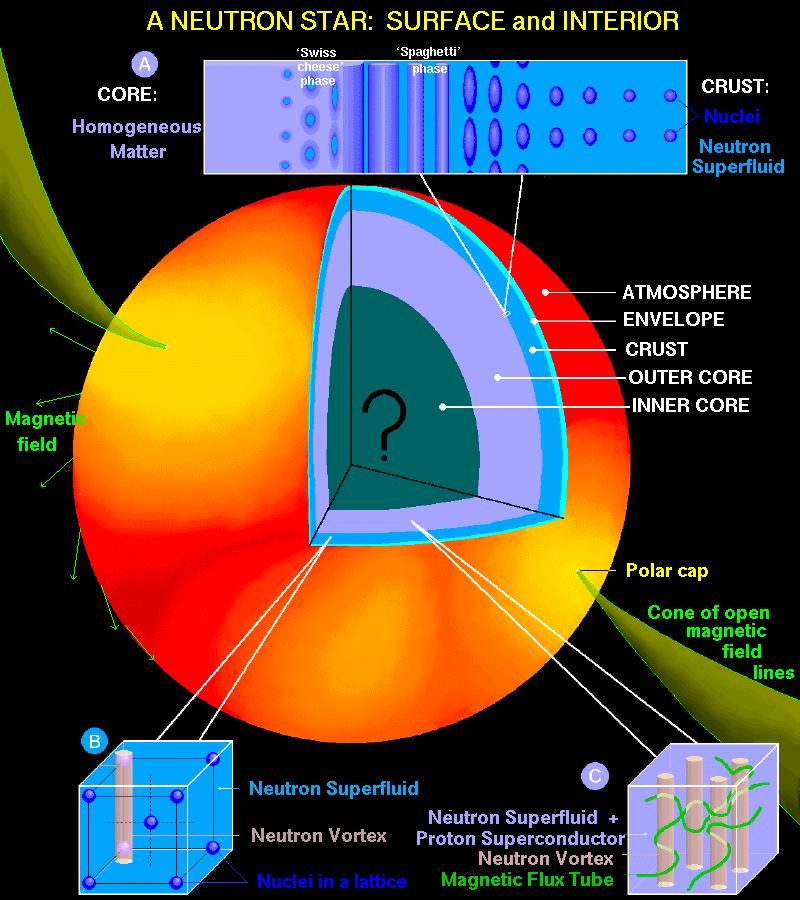}
   \hspace{0.1cm}
  \includegraphics[height=2.25in]{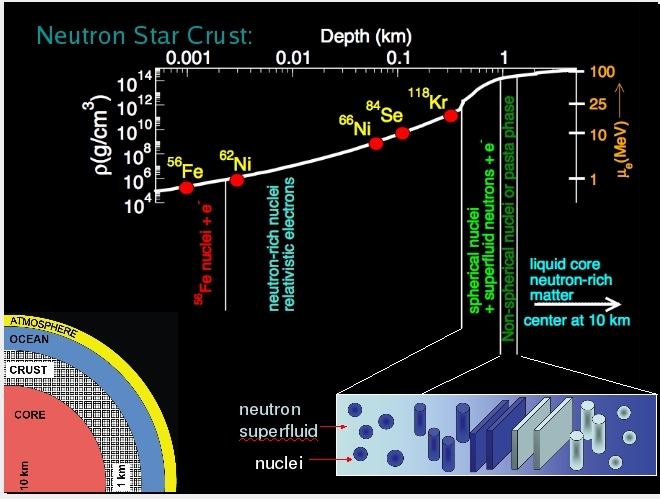}
  \vspace{-0.2cm}
  \caption{Two physically accurate renditions of the structure, composition, 
  	       and possible phases of a neutron star. Courtesy of Dany Page 
	       (left panel) and Sanjay Reddy (right panel).}
 \label{Fig1}
\end{center}
\end{figure}

In the next few sections we embark on a journey through a neutron 
star. According to Baade and Zwicky, the most common perception 
of a neutron star is that of a uniform assembly of extremely closed 
packed neutrons\,\cite{Baade:1934}. We will now show, however, 
how the reality is far different and much more interesting. In 
particular, as we journey through the neutron star we will discover a 
myriad of exotic states of matter and will discuss the critical role that 
laboratory experiments can play in elucidating their fascinating nature. 
Because of their enormous relevance to nuclear physics, we focus our 
attention on three components of the neutron star: (a) the outer crust, 
(b) the inner crust, and (c) the outer core. For two recent reviews on the
exotic nature of the stellar crust see\,\cite{Chamel:2008ca,Bertulani:2012}  
and references therein. The three regions are clearly highlighted in 
Fig.\,\ref{Fig1}, which includes two physically accurate renditions of a 
neutron star. In particular, the question mark at the center of the 
left-hand illustration denotes the possibility that the inner stellar core 
harbors exotic states of matter, such as hyperons, meson condensates, 
and strange quark matter. Although this idea is enormously provocative, 
at present there is simply not enough experimental information to properly 
constrain the dynamics of the inner core. Thus, the possible existence 
of such exotic states of matter will be ignored hereafter. Although the 
stellar atmosphere and the envelope will also be ignored, we briefly 
discuss now some of its most relevant features.

Because of the enormous gravitational fields around a neutron star, the 
atmosphere is believed to be about a mere 10\,cm thick. However, the 
atmosphere shapes the thermal radiation from the photosphere which is 
customarily assumed to be that of a black body. Hence, detailed 
knowledge of the atmosphere is critical for the reliable extraction of, 
for example, stellar radii. In turn, the 100\,m envelope acts as a blanket 
that modulates the huge temperature gradient between the core and the 
crust. For more information about the role of the stellar atmosphere and 
envelope see Ref.\,\cite{Page:2006ud} and references contained therein.

\subsection{The Outer Crust}
\label{OuterCrust}

The outer crust of a neutron star comprises a region that spans 
almost 7 orders of magnitude in baryon density;  from about 
$10^{4}{\rm g/cm^{3}}$ to $4\times 
10^{11}{\rm g/cm^{3}}$\,\cite{Baym:1971pw}. Recall that
nuclear-matter saturation density is 
$\rhozero\!\simeq\!2.4\,\times10^{14}{\rm g/cm^{3}}$, which 
corresponds to a baryon density of $\nzero\!\simeq\!0.15\,{\rm fm}^{-3}$.
Thus, these densities are significantly lower that those encountered in 
the interior of the atomic nucleus. However, at these densities the 
electrons---which represent a critical component of the star in order 
to maintain the overall charge neutrality of the system---have been 
pressure ionized. Thus, they can be accurately described by a 
relativistic free Fermi gas. Moreover, given that at these densities 
the average inter-nucleon distance is significantly larger than the 
range of the strong nuclear force, the uniform ground state becomes 
unstable against cluster formation. That is, under these conditions 
of density it is energetically favorably for translational invariance to 
be broken and for the individual nucleons to cluster into ``normal'' 
nuclei. Thus, the outer stellar crust consists of isolated nuclei 
embedded in a uniform electron gas. Moreover, because the 
short-range nuclear force saturates within the individual clusters, 
nuclei interact exclusively via the long-range Coulomb interaction. 
This promotes the formation of a Coulomb crystal of
nuclei arranged in a body-centered-cubic (bcc) lattice that itself
is embedded in a neutralized uniform electron gas\,\cite{Baym:1971pw}.
In the particular case of the top layers of the crust where the density is 
at its lowest, the energetically preferred nucleus is ${}^{56}$Fe; see
the right-hand panel of Fig.\,\ref{Fig1}.  Recall that ${}^{56}$Fe is the
nucleus with the lowest mass per nucleon. 

However, as one moves inward towards the center of the star, the density 
increases, and so does the electronic contribution to the total energy. Thus, 
it becomes energetically advantegeous to remove a fraction of the electrons 
(through electron capture) albeit at the expense of an increase in the 
neutron-proton asymmetry. In this manner, the energetically most favorable 
nucleus emerges from a competition between the electronic contribution, 
which favors a small electron (and proton) fraction, and the nuclear symmetry 
energy which, in turn, favors symmetric nuclei. Hence, the nuclear contribution
to the composition of the outer crust appears in the form of a nuclear mass
table that is generated from a combination of experimental data and theoretical 
predictions. Indeed, in full thermodynamic equilibrium, one determines the 
crustal composition by minimizing the chemical potential of the system ($\mu$) 
at zero temperature and fixed pressure. That  is\,\cite{RocaMaza:2008ja},
\begin{equation}
 \mu(A,Z;P) =\frac{M(N,Z)}{A}+\frac{Z}{A}\mu_{e}
                   -\frac{4}{3}C_{\ell}\frac{Z^{2}}{A^{4/3}}\pFermi \,.
 \label{ChemPotential}
\end{equation}
The total chemical potential consists of nuclear, electronic, and lattice
contributions. As already mentioned, computing the nuclear contribution 
requires of ``only'' a reliable nuclear mass table. Moreover, given that at
the relevant densities the electrons can be accurately modeled by 
a Fermi-gas distribution, its contribution to the chemical potential
is simply given by
\begin{equation}
 \mu_{e} = \sqrt{(\pFermi^{e})^{2}+m_{e}^{2}} 
             = \sqrt{(y\pFermi)^{2}+m_{e}^{2}} \,,
 \label{EChemPotential}
\end{equation}
where $y\!=\!Z/A$ is the electron fraction and the Fermi momentum
$\pFermi$ is related to the baryon density $n$ by 
\begin{equation}
 \pFermi = \left(3\pi^2 n\right)^{1/3} \,.
 \label{pFermi}
\end{equation}
Finally, the last term in Eq.\,(\ref{ChemPotential}) represents the 
complicated lattice contribution. This contribution is complex
because it involves the long-range nature of the Coulomb interaction.
Nevertheless, the overall charge neutrality of the system ensures 
that the contribution is finite\,\cite{Coldwell:1960,Sholl:1967}. In
the particular case of the energetically preferred bcc lattice, one
obtains\,\cite{Baym:1971pw}
\begin{equation}
 \varepsilon_{\ell}(A,Z;n)=-
 C_{\ell} \frac{Z^{2}}{A^{4/3}}p^{}_{\rm F} \,,
 \label{EoverALatt2}
\end{equation}
where $C_{\ell}\!=\!3.40665\!\times\!10^{-3}$ is a dimensionless 
constant\,\cite{RocaMaza:2008ja}.
Note that although the expression for the chemical potential is 
written in terms of the density rather than the pressure, chemical 
equilibrium demands that the minimization of $ \mu(A,Z;P)$ be 
carried out at a fixed pressure rather than at a fixed density. 
Thus, one needs an equation of state to properly relate them. 
However, given that for the outer crust the pressure is dominated 
by the degenerate electrons (with only a small lattice contribution) 
to a very good approximation the relevant equation of state is that 
of a relativistic Fermi gas of electrons.
\vspace{-0.1cm}
\begin{figure}[h]
\begin{center}
  \includegraphics[height=7cm]{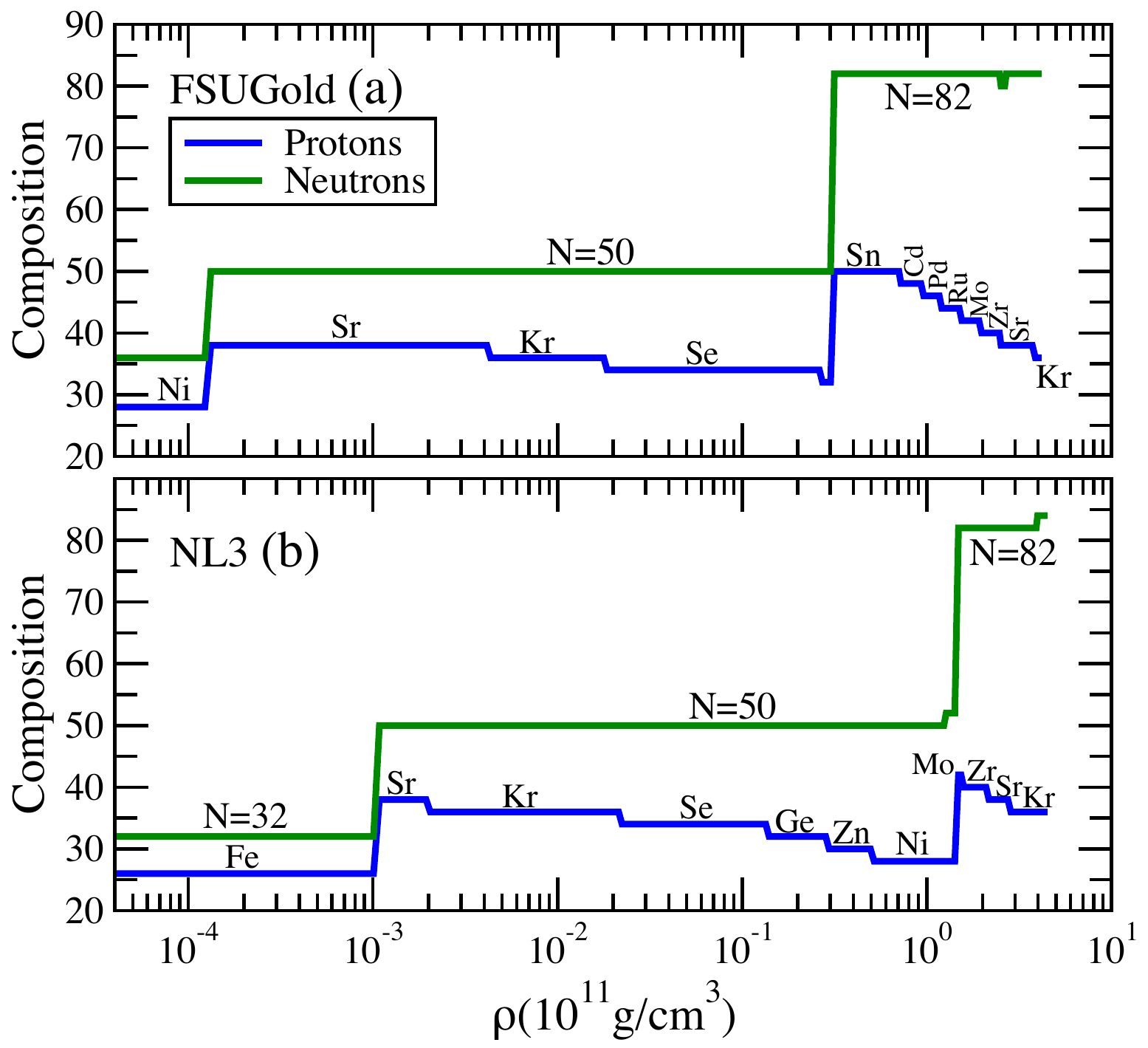}
   \hspace{0.1cm}
  \includegraphics[height=7cm]{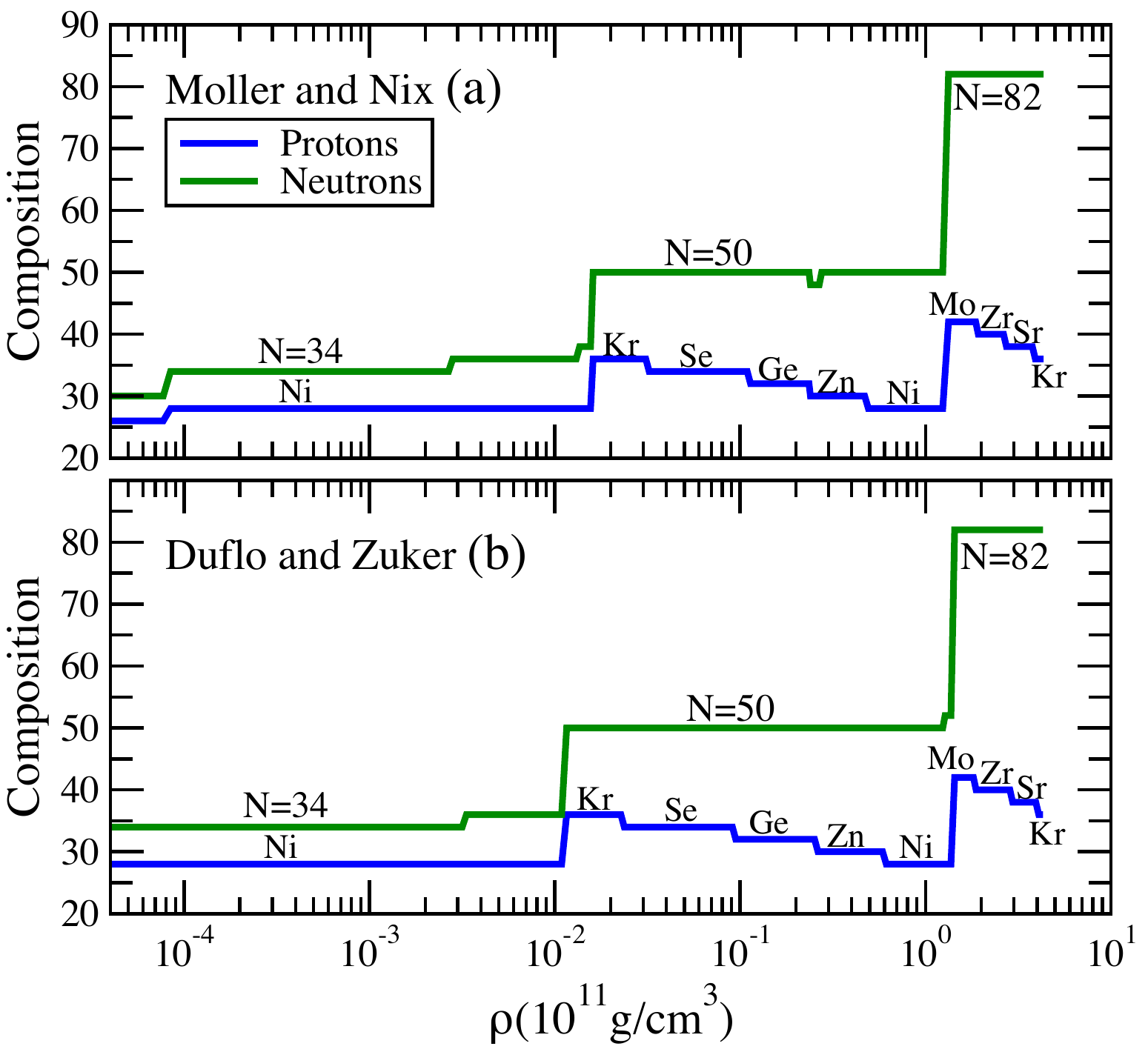}
  \vspace{-0.2cm}
  \caption{Composition of the outer crust as predicted by two
               accurately-calibrated relativistic mean field models (left) and
               by two microscopic-macroscopic models (right). Protons are 
               depicted with the blue (lower) line whereas neutrons with the 
               green (upper) line.}
 \label{Fig2}
\end{center}
\end{figure}
\vspace{-0.2cm}

To assess the sensitivity of the crustal composition to the mass
model we display in Fig.\,\ref{Fig2} predictions using two accurately
calibrated relativistic EDFs, i.e.,
FSUGold\,\cite{Todd-Rutel:2005fa} and
NL3\,\cite{Lalazissis:1996rd,Lalazissis:1999} as well as two highly
successful microscopic-macroscopic (mic-mac) models; one due to
M\"oller and collaborators\,\cite{Moller:1993ed,Moller:1997bz} and 
the other one due to Duflo and
Zuker\,\cite{Duflo:1994,Zuker:1994,Duflo:1995}.  As already
anticipated, the top (i.e., most dilute) layers of the outer crust
consist of a crystalline lattice of ${}^{56}$Fe nuclei embedded in a
uniform electron gas; for a more aesthetic view see the right-hand 
panel of Fig.\,\ref{Fig1}. However, as the density
increases---and with it the electron chemical potential---it becomes
energetically favorable to replace ${}^{56}$Fe with a slightly more
neutron-rich nucleus. That is, at slightly
higher densities ${}^{56}$Fe ceases to be the most stable nucleus. 
Rather, the slightly more isospin asymmetric (even-even) Nickel 
isotopes become energetically favored. As the density continues to 
increase further, the system must decide whether to reduce $Z$ at
neutron number $N\!\approx\!34$ or to increase both $N$ and $Z$ 
by jumping to the next magic shell at $N\!=\!50$. Although it is seen 
that all four models favor an eventual jump to the $N\!=\!50$ shell, 
the predictions for the density at which the jump occurs is highly 
model dependent. Indeed, whether FSUGold predicts the transition 
at a density of about $10^{7}{\rm g/cm^{3}}$, both mic-mac models 
suggest a density that is almost two orders of magnitude larger. Our 
results also indicate that in all cases the proton fraction decreases
systematically with increasing density in an effort to reduce the
electronic contribution to the chemical potential. Yet, at one point
reducing the electron fraction even further becomes too expensive 
for the symmetry energy to sustain and the system jumps to the 
next plateau at magic number $N\!=\!82$. We want to underscore 
that both the transition density as well as the crustal composition 
depend sensitively on the symmetry energy at sub-saturation density. 

In the particular case of the two microscopic models (FSUGold and 
NL3) the symmetry energy may be computed at all densities, so a 
study of the trends displayed in Fig.\,\ref{Fig2} are illuminating. At 
a density of relevance to finite nuclei, which consists of an average
between the nuclear interior and the nuclear surface, the symmetry
energy is known to be larger for FSUGold than for NL3. This implies 
that a neutron-proton mismatch at those densities is more costly
for FSUGold than for NL3\,\cite{Todd:2003xs}. Thus, the tolerance 
to a larger neutron-proton asymmetry is responsible for delaying 
the transition to the next higher plateau for NL3 relative to FSUGold,
a fact that is clearly evident in the figure. By the same token, NL3 
predicts a more exotic crustal composition than FSUGold. Indeed,
whereas FSUGold suggests the formation of ${}^{132}$Sn 
at the transition to the $N\!=\!82$ shell, NL3 predicts the formation 
of the significantly more neutron-rich isotope ${}^{124}$Mo. Note 
that as the density continues to increase even further, the 
neutron-proton asymmetry will become so large that the neutron
drip line will be reached. For all models, the drip line is predicted to 
occur at a density of about $4\!\times\!10^{11}{\rm g/cm^{3}}$ and
with the formation of the highly exotic ${}^{118}$Kr isotope.

We close this section with a brief comment on the assumptions and
extrapolations required to predict the composition of the outer crust.
As just described, there are three main regions in the nuclear chart
that are of direct relevance to the outer crust: (a) The Fe-Ni region,
(b) the $N\!=\!50$ isotones from Ni ($Z\!=\!28$) to Sr ($Z\!=\!38$), 
and (c) the $N\!=\!82$-isotone region from Kr ($Z\!=\!36$) to Sn 
($Z\!=\!50$). In regards to (a), all the masses in this region have 
been measured with great precision\,\cite{AME:2012}. In the case 
of (b), precise mass values exist, but only for the cases of ${}^{88}$Sr, 
${}^{86}$Kr, ${}^{84}$Se, ${}^{82}$Ge, and ${}^{80}$Zn, but not for 
either ${}^{78}$Ni or ${}^{82}$Zn\,\cite{AME:2012,Goriely:2010bm}. 
However, in a pioneering Penning trap experiment with the ISOLTRAP 
setup at the ISOLDE-CERN facility, the mass of ${}^{82}$Zn has been 
recently determined\,\cite{Wolf:2013ge}. This new mass determination
has ruled out the presence of ${}^{82}$Zn in the outer crust and provides
the most stringent constraint to date on its composition profile. Finally,
the $N\!=\!82$-isotone region remains largely unexplored and it is likely
to remain so even after the construction of a new generation of rare
isotope facilities. Thus, the only hope to elucidate the composition of 
the bottom layers of the outer crust is through theoretical modeling.
In this regard, measuring a large number of as yet unknown masses 
of exotic nuclei---even if of no direct relevance to the composition 
of the outer crust---will still be instrumental in guiding the calibration 
of future nuclear density functionals.

\subsection{The Inner Crust}
\label{Inner Crust}
The inner crust of a neutron star comprises the region from
neutron-drip density up to the density at which uniformity in 
the system is restored; about one third to one half of normal 
nuclear density. However, the transition from the highly 
ordered Coulomb crystal to the uniform liquid is both 
interesting and complex. This is because distance scales 
that are well separated in both the crystalline phase, where 
the long-range Coulomb interaction dominates, and in the 
uniform phase, where the short-range strong 
interaction dominates, become comparable in the inner
stellar crust. This unique situation involving competing 
distance scales gives rise to \emph{Coulomb frustration}.  
Frustration, a universal phenomenon characterized by the 
existence of a very large number of low-energy configurations,
emerges from the impossibility to simultaneously minimize all 
elementary interactions in the system. Ultimately, the competition 
between the short-range nuclear interaction and the long-range
Coulomb repulsion results in the formation of complex topological 
structures collectively referred to as \emph{nuclear pasta}. Given 
that these complex structures are very close in energy, it has been 
speculated that the transition from the highly ordered crystal 
to the uniform phase must proceed through a 
series of changes in the dimensionality and topology of these
structures\,\cite{Ravenhall:1983uh,Hashimoto:1984}. Moreover, 
due to the preponderance of low-energy states, frustrated 
systems display an interesting and unique low-energy dynamics.
We note that a seemingly unrelated condensed-matter problem,
namely, the strongly-correlated electron gas, also displays 
Coulomb frustration. In the case of the electron gas, one 
aims to characterize the transition from the low-density Wigner 
crystal, where the long-range Coulomb potential dominates, to 
the uniform Fermi liquid, where the kinetic energy 
dominates\,\cite{Fetter:1971}. It has been shown that such a
transition must be mediated by the emergence of ``microemulsions'', 
namely, exotic pasta-like structures with interesting topologies. Indeed, 
it has been proven that in two-spatial dimensions a direct first-order phase 
transition is forbidden in the presence of long-range (e.g., Coulomb)
forces~\cite{Jamei:2005}. 

\begin{figure}[h]
\begin{center}
 \includegraphics[height=7cm]{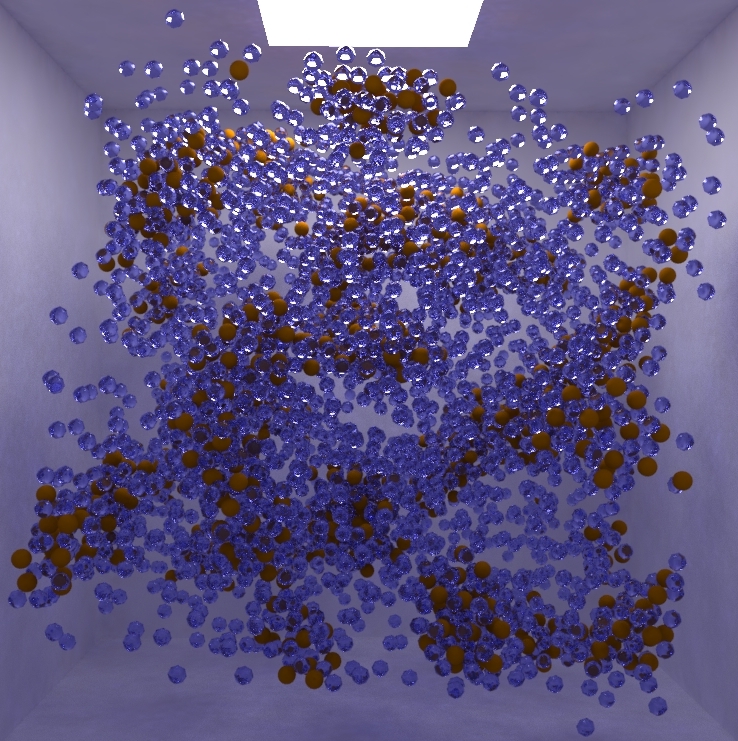}
 \includegraphics[height=7cm]{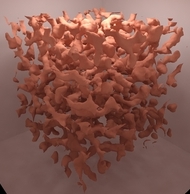} 
 \vspace{-0.2cm}
   \caption{(Left panel) A snapshot of a Monte Carlo simulation 
    for one configuration of 4,000 nucleons at a baryon density 
    of $0.025\,{\rm fm}^{-3}$, a proton fraction of $Z/A\!=\!0.2$, 
    and a temperature of 1\,MeV. (Right panel) The 
    $0.03\,{\rm fm}^{-3}$ proton density iso-surface for one 
    configuration of 100,000 nucleons at a density of $0.05\,{\rm fm}^{-3}$, 
    a proton fraction of $Z/A\!=\!0.2$, and a temperature of 1\,MeV. 
    Note that the simulation volume in this latter case is fairly large, 
    i.e., $V\!=\!(126\,{\rm fm})^{3}$.}
 \label{Fig3}
\end{center} 
\end{figure}

To illustrate the complexity of the pasta phase we display in Fig.\,\ref{Fig3} 
two snapshots obtained from Monte-Carlo and Molecular-Dynamics simulations 
of a nuclear system at densities of relevance to the inner stellar
crust\,\cite{Horowitz:2004yf,Horowitz:2004pv}. The left-hand panel displays 
how at a density of about one sixth of normal nuclear density and a proton
fraction of $Z/A\!=\!0.2$, the system organizes itself into neutron-rich clusters 
(i.e., ``nuclei'') of complex topologies that are surrounded by a dilute vapor of 
likely superfluid neutrons. In turn, the right-hand panel displays the complex 
topology of a proton iso-surface (i.e., a surface of constant proton density) for
a system of 100,000 nucleons. Such complex pasta 
structures may have a significant impact on various transport properties, 
such as neutrino propagation and electron conductivity. We should underscore
that the emergence of such complex structures is a true dynamical effect 
associated with Coulomb frustration, as no a-priori shapes (such as spheres, 
rods, slabs, etc.) are ever assumed.

A great advantage of molecular-dynamics (MD) simulations is that many-body 
correlations are properly accounted within the 
formalism\,\cite{Horowitz:2004yf,Horowitz:2004pv,Horowitz:2005zb,
Watanabe:2003xu,Watanabe:2004tr, Watanabe:2009vi}. However, given 
their classical nature, MD simulations fail to capture any quantum-mechanical 
detail that the system may be sensitive to; for example, the superfluid nature
of the dilute neutron vapor. In contrast, mean-field approximations incorporate 
quantum-mechanical effects (at least on average) but fail to describe the 
important and complicated clustering 
correlations\,\cite{Maruyama:2005vb,Avancini:2008zz,Avancini:2008kg,
Newton:2009zz,Shen:2011kr}.  Nevertheless, because the robustness of 
Coulomb frustration, both set of theoretical approaches seem to reach 
similar conclusions. For example, for relatively large proton fractions in 
the $0.3$--$0.5$ range (such as in core-collapse supernovae) there 
appears to be general agreement that the transition from the ordered 
Coulomb crystal to the uniform phase must proceed via intermediate pasta 
phases. What is unclear, however, is whether such exotic pasta shapes 
can develop in the proton-poor environment characteristic of the inner 
stellar crust\,\cite{Piekarewicz:2011qc}. Note that mean-field models 
that impose $\beta$-equilibrium predict proton fractions at densities of 
relevance to the inner crust of only a few 
percent\,\cite{Maruyama:2005vb,Avancini:2008zz,Avancini:2008kg,Shen:2011kr}.

Another enormous challenge associated with the purported pasta phases
is the identification of a set of astrophysical observables that are sensitive 
to their formation. Colloquially, we refer to this challenge as ``how to smell
the pasta?'' A possible manifestation of the nuclear pasta on the dynamics
of neutron stars has been suggested recently by Pons, Vigan\`o, and 
Rea\,\cite{Pons:2013nea}. These authors have identified a special class
of rotation-powered pulsars---\emph{the isolated x-ray pulsars}---that 
appear to display spin periods shorter than about 12 seconds. Although
enormously stable, rotation-powered pulsars are known to slow down,
albeit very slowly, due to the emission of magnetic dipole radiation. The
lack of isolated x-ray pulsars with spin periods longer than 12 seconds
seems to suggest magnetic field decay due to the existence of a highly 
resistive layer in the inner crust; such a layer has been speculated to be 
the exotic nuclear pasta phase\,\cite{Pons:2013nea}. Very recently, Horowitz 
and collaborators have carried out large molecular dynamics simulations to 
explore a possible magnetic field decay in the inner crust due to the
existence of a nuclear pasta phase\,\cite{Horowitz:2014xca}. In 
particular, it was concluded that the formation of \emph{topological
defects} in the nuclear pasta could reduce both its electrical and 
thermal conductivity. Hence, the formation of this ``highly resistive
layer'' could promote magnetic field decay and may ultimately explain
the lack of x-ray pulsars with long spin periods\,\cite{Pons:2013nea}.

\subsection{The Outer Core}
\label{OuterCore}

Structurally, the stellar core is by far the most critical component of the star. Because 
of the enormous interior densities, practically all the mass and most of the size reside 
in the stellar core. At densities of about $10^{14}{\rm g/cm^{3}}$, the pasta phase 
``melts'' and uniformity in the system is restored.  It is in the stellar core where the 
original perception of Baade and Zwicky\,\cite{Baade:1934} is finally realized, namely, 
a neutron star as a uniform assembly of extremely closed packed neutrons. However, 
in order to maintain both chemical equilibrium and charge neutrality, a small fraction (of 
about 10\%) of protons and leptons is also required. Remarkably, given that the 
densities in the stellar core are so large, the leptonic component consists of both 
electrons and \emph{muons}; indeed, a neutron star typically contains about 
$10^{56}$ muons! Although exotic, the presence of muons is a model-independent
consequence of chemical equilibrium. Instead, the physics of some of the more exotic 
states of matter that have been speculated to exist in the stellar core---such as hyperons, 
meson condensates, and quark matter---are much more uncertain. Thus, we limit ourselves 
to model the stellar core exclusively in terms of non-exotic constituents, namely, neutrons, 
protons, electrons, and muons. Moreover, this approach enables us to test and improve 
relativistic density functionals that were calibrated using only the properties of finite 
nuclei. This is highly desirable as the equation of state of dense neutron-rich matter 
is poorly constrained by laboratory observables. 

The cleanest constraint on the EOS at high density emerges from the accurate 
measurement of massive neutron stars. In this regard, enormous progress 
has been made by the recent observations of two massive neutron stars by 
Demorest\,\cite{Demorest:2010bx} and Antoniadis\,\cite{Antoniadis:2013pzd}
at the Green Bank Telescope. In particular, the accurate mass measurement of 
PSR J1614-2230 (with a  mass of $M_{\star}\!=\!1.97\!\pm\!0.04M_{\odot})$ 
was carried out using ``Shapiro delay", a beautiful and powerful technique that 
relies on the tiny delay in the arrival of the pulsar signal as it ``deeps down'' into 
the gravitational well of its white-dwarf companion\,\cite{Demorest:2010bx}. This 
observation provides a clear example of the synergy between nuclear physics,
astrophysics, and general relativity. 

\begin{figure}[h]
\begin{center}
 \includegraphics[height=6.50cm]{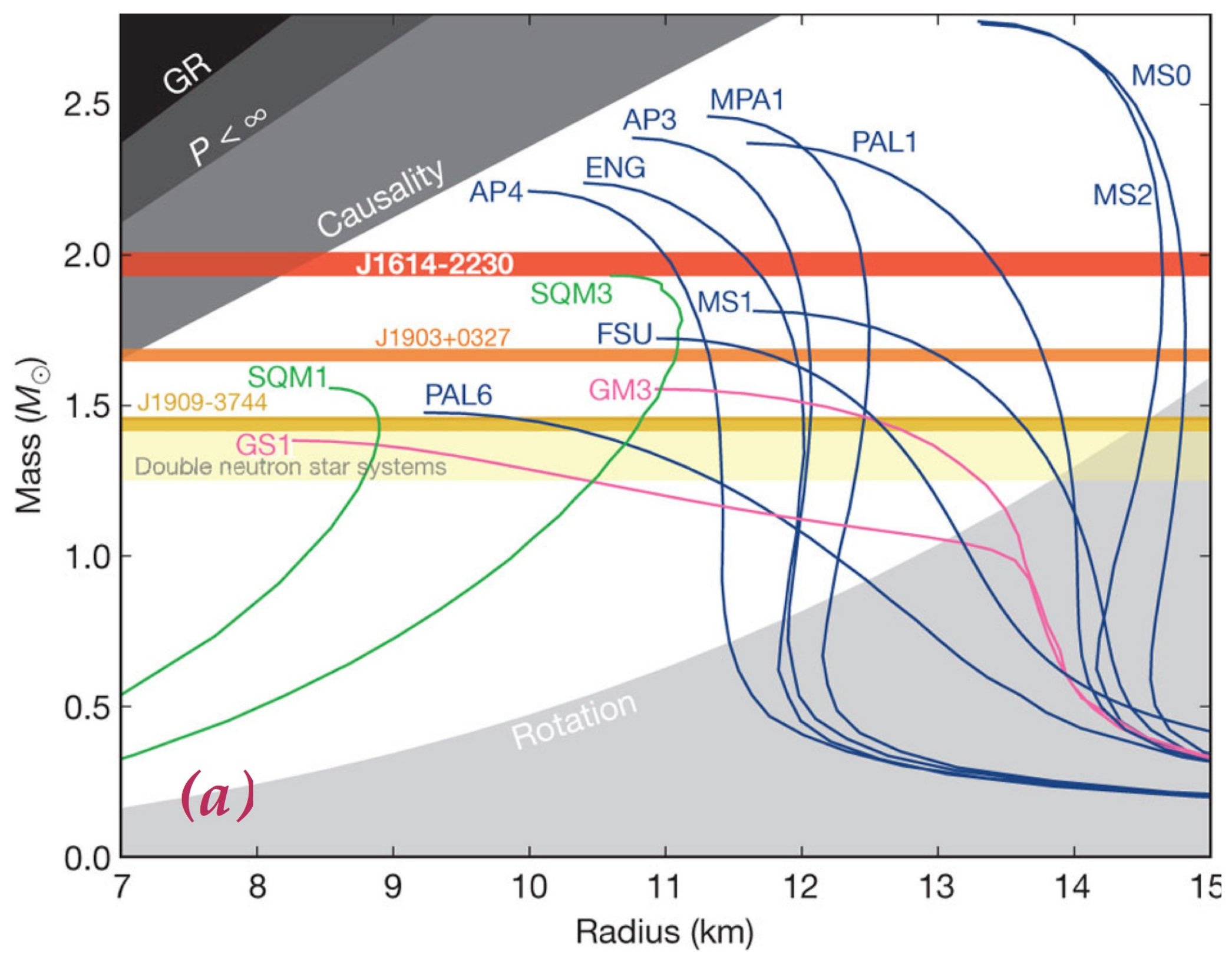}
 \includegraphics[height=6.60cm]{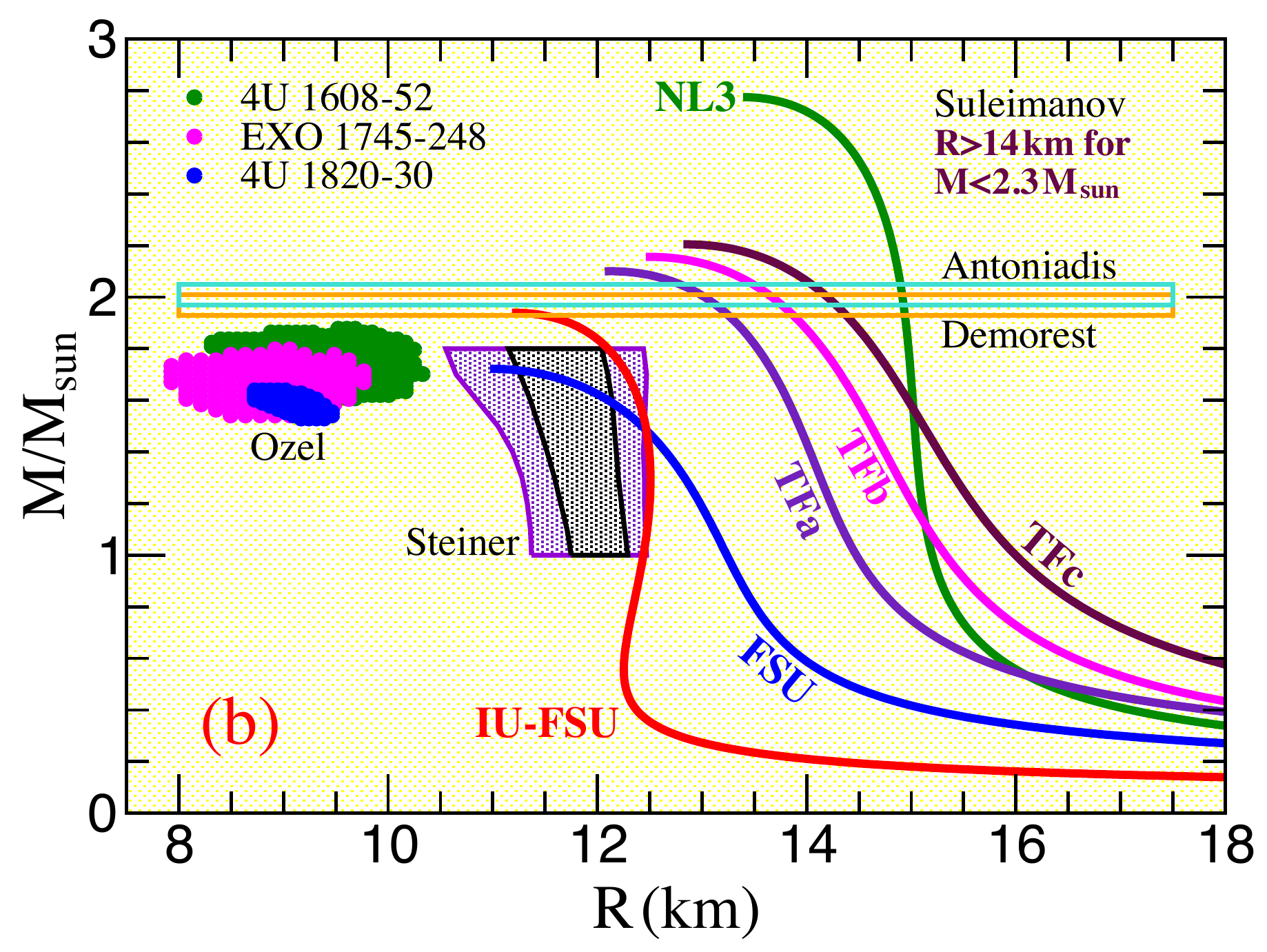} 
 \vspace{-0.2cm}
   \caption{(Left panel) Predictions for the mass-vs-radius relation for a variety of models 
                of the EOS using both exotic and non-exotic cores. The recent discovery of 
                the massive neutron star PSR J1614-2230 by Demorest and collaborators is
                clearly indicated in the figure\,\cite{Demorest:2010bx}. (b) (Right panel) 
                Constraints on both stellar masses and radii extracted from various analyses 
                of x-ray bursts\,\cite{Ozel:2010fw,Steiner:2010fz,Suleimanov:2010th}. Also 
                shown are constraints obtained from the measurement of two massive neutron 
                stars by Demorest\,\cite{Demorest:2010bx} and 
                Antoniadis\,\cite{Antoniadis:2013pzd}.}
 \label{Fig4}
\end{center} 
\end{figure}

The impact of such mass measurements 
on the predictions of the \emph{mass-vs-radius} relation of neutron stars
is displayed in Fig.\,\ref{Fig4}a. Note that the constraint from PSR J1614-2230 
alone can rule out models with equations of state that are too soft to support a 
$2M_{\odot}$ neutron star, such as those that predict the existence of exotic 
cores. Undoubtedly, the quest for even more massive neutron stars will continue 
and this will provide unique and valuable insights into the behavior of ultra-dense 
matter.

Although enormously valuable, the accurate measurement of massive neutron stars 
does not impose any significant constraint on stellar radii. Indeed, Fig.\,\ref{Fig4}a 
indicates that predictions for the radius of a ``canonical'' $1.4M_{\odot}$ neutron star 
can vary by almost a factor of two; between 8\,km and 15\,km. Whereas observations 
of various spectroscopic phenomena in x-ray bursters are promising in the quest to
measure stellar radii, the approach 
currently suffers from large systematic uncertainties. For example, in one of the first 
analyses of this kind by \"Ozel, Baym, and G\"uver\,\cite{Ozel:2010fw}, it was 
suggested that neutron stars with masses of about $1.4M_{\odot}$ have very small
radii of about $8$-$\!10$\,km; see Fig.\,\ref{Fig4}b. This conclusion seems to favor
models with exotic cores\,\cite{Fattoyev:2010rx} which, in turn, are disfavored by the 
observation of $2M_{\odot}$ neutron stars. However, it was recognized soon after,
first by Steiner, Lattimer, and Brown\,\cite{Steiner:2010fz} and shortly after by 
Suleimanov\,\cite{Suleimanov:2010th}, that if one corrects for systematic 
uncertainties in the analysis by \"Ozel, then it is possible to obtain larger (in 
the case of the former) or even much larger (in the case of the latter) stellar radii. 
Indeed, the results by Steiner\,\cite{Steiner:2010fz} depicted in Fig.\,\ref{Fig4}b 
by the two shaded areas that indicate 1$\sigma$ and 2$\sigma$ contours, suggest 
stellar radii in the $10$-$\!13$ km range. However, even this more conservative
limit has been challenged by Suleimanov, who has proposed a \emph{lower 
limit} on the stellar radius of 14\,km for neutron stars with masses below 
2.3M$_{\odot}$\,\cite{Suleimanov:2010th}. This conclusion suggests that the 
neutron-star matter EOS must be stiff, in agreement with the recent measurement
of massive stars. 

While at present a reliable technique based on thermal 
emissions from x-ray bursts has not yet been realized, a recent analysis limited
to the observation of accreting neutron stars during quiescence, the so-called 
\emph{quiescent low-mass x-ray binaries} (qLMXBs), has challenged our current 
understanding of the equation of state of dense matter. qLMXBs are particularly
attractive as they appear to be free from some of the uncertainties that plague x-ray 
bursters. By analyzing the 
thermal spectra of five qLMXBs inside globular clusters, Guillot and collaborators 
have reported a \emph{common radius} for all five sources of 
$R_{NS}\!=\!9.1^{+1.3}_{-1.5}$\,km at a 90\% confidence level\,\cite{Guillot:2013wu}. 
One should mention that whereas the approach developed in Ref.\,\cite{Guillot:2013wu} 
provides a careful accounting of all uncertainties, some of the assumptions---such as 
the single common radius---and some of the adopted uncertainties have been called into 
question\,\cite{Lattimer:2013hma}. Regardless, it is interesting to note that few, if
any, nuclear density functionals---whereas relativistic or non-relativistic---can accommodate 
both a small stellar radius and a large limiting mass\,\cite{Guillot:2013wu}. However, we are 
confident that with new space missions---such as GAIA---that will provide unprecedented 
positional measurements, many of the current problems plaguing such challenging analyses 
will be mitigated.
\begin{figure}[h]
 \begin{center}
 \includegraphics[height=8cm]{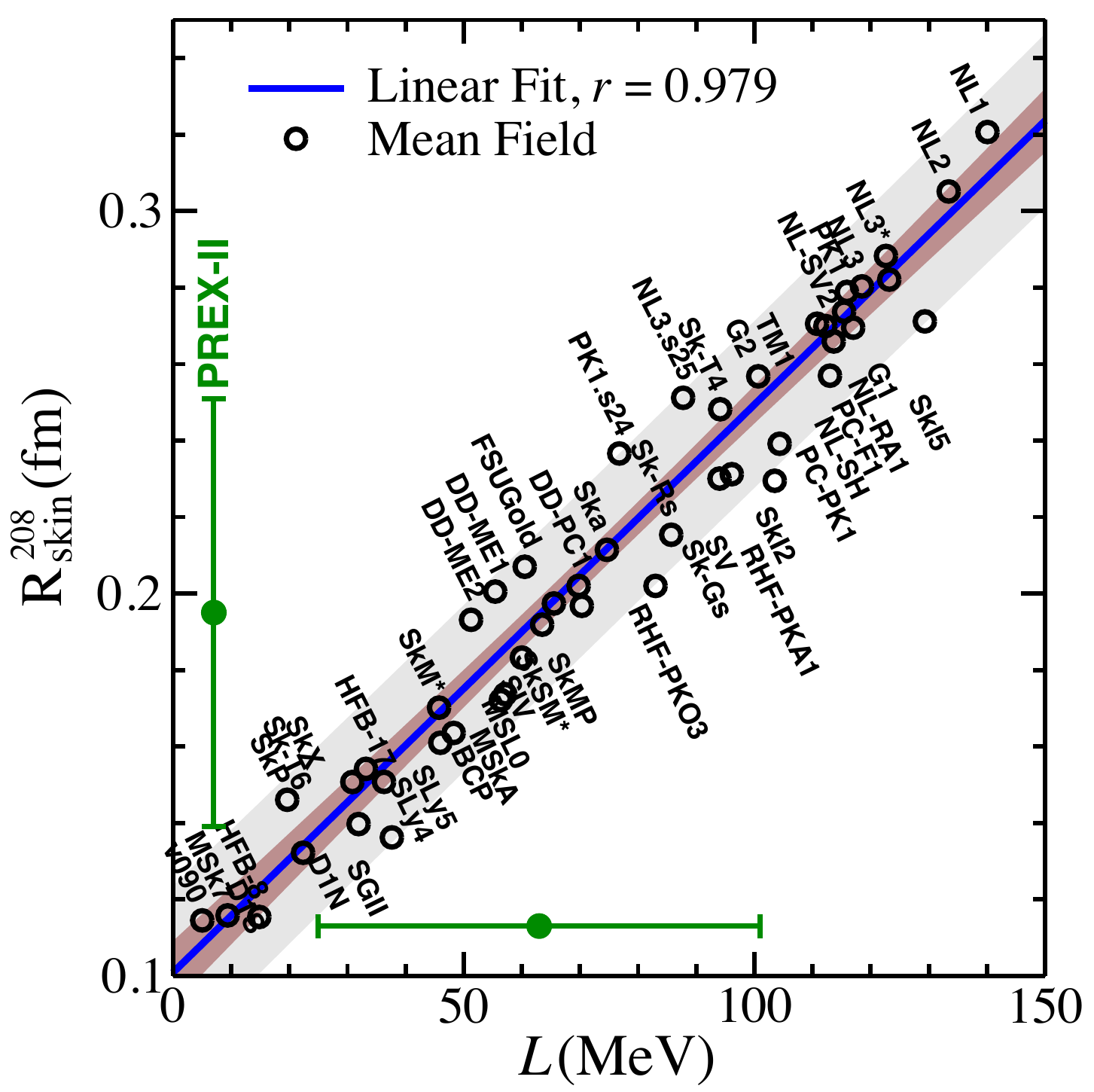}
 \vspace{-0.2cm}
  \caption{Predictions from a large number of nuclear density functionals for the 
  	       neutron-skin thickness of ${}^{208}$Pb and the slope of the symmetry
	       energy $L$\,\cite{RocaMaza:2011pm}. Constraints from an anticipated
	       upgraded PREX measurement (``PREX-II'') have been incorporated 
	       into the plot.} 
 \label{Fig5}
 \end{center}
\end{figure}

Whereas laboratory experiments are of marginal utility in constraining the limiting mass 
of a neutron star, they play an essential role in constraining stellar radii. This is because 
the radius of a neutron star is sensitive to the density dependence of the symmetry energy 
in the immediate vicinity of nuclear-matter saturation density\,\cite{Lattimer:2006xb}. A 
fundamental property of the EOS that has received considerable  attention over the last 
decade is the slope of the symmetry energy at saturation density\,\cite{Piekarewicz:2008nh}. 
As already shown in Eq.\,(\ref{PvsL}), the slope of the symmetry energy $L$ is proportional
to the pressure of pure neutron matter at saturation density. In turn, the slope of the symmetry
energy is also strongly correlated to a myriad of neutron-star observables\,\cite{Horowitz:2000xj,
Horowitz:2001ya,Horowitz:2002mb,Fattoyev:2012rm}. 
Remarkably, $L$ is also strongly correlated to the thickness of the neutron skin of heavy 
nuclei\,\cite{Brown:2000,Furnstahl:2001un}, which is defined as the difference between the 
neutron and proton root-mean-square radii. The physical reason behind this correlation is 
particularly insightful. Heavy nuclei favor a neutron excess as a result of the repulsive 
Coulomb interaction between the protons. Energetically, it is advantageous---to both 
the surface tension and to the symmetry energy---to form an isospin symmetric 
($N\!=\!Z$) core. So the basic question to be answered is \emph{where do the extra 
neutrons go?} Placing them in the core reduces the surface tension but increases 
the symmetry energy. Moving the excess
neutrons to the surfaces increases the surface tension but reduces the symmetry 
energy, which is lower in the dilute surface than in the dense core. So the thickness 
of the neutron skin emerges from a competition between the surface tension and 
the \emph{difference} between the value of the symmetry energy at the surface 
relative to that at the center; this difference is encoded in the slope of the symmetry
energy $L$\,\cite{Horowitz:2014bja}. Hence, if $L$ is large, then it is energetically
advantegeous to move the neutrons to the surface, which results in a thick neutron 
skin. This suggests a powerful correlation: \emph{the larger the value of $L$ the 
thicker the neutron skin}\,\cite{Horowitz:2001ya}. The strong correlation between 
$L$ and the neutron-skin thickness of ${}^{208}$Pb ($R_{\rm skin}^{208}$) has 
been verified using a large and representative set of density functionals---both 
relativistic and non-relativistic---and is displayed in 
Fig.\,\ref{Fig5}\,\cite{RocaMaza:2011pm}. The strong correlation coefficient of 
$r\!=\!0.979$ suggests how a laboratory observable such as $R_{\rm skin}^{208}$ 
may serve to determine a fundamental parameter of the equation of state.

\begin{figure}[h]
 \begin{center}
 \includegraphics[height=8cm]{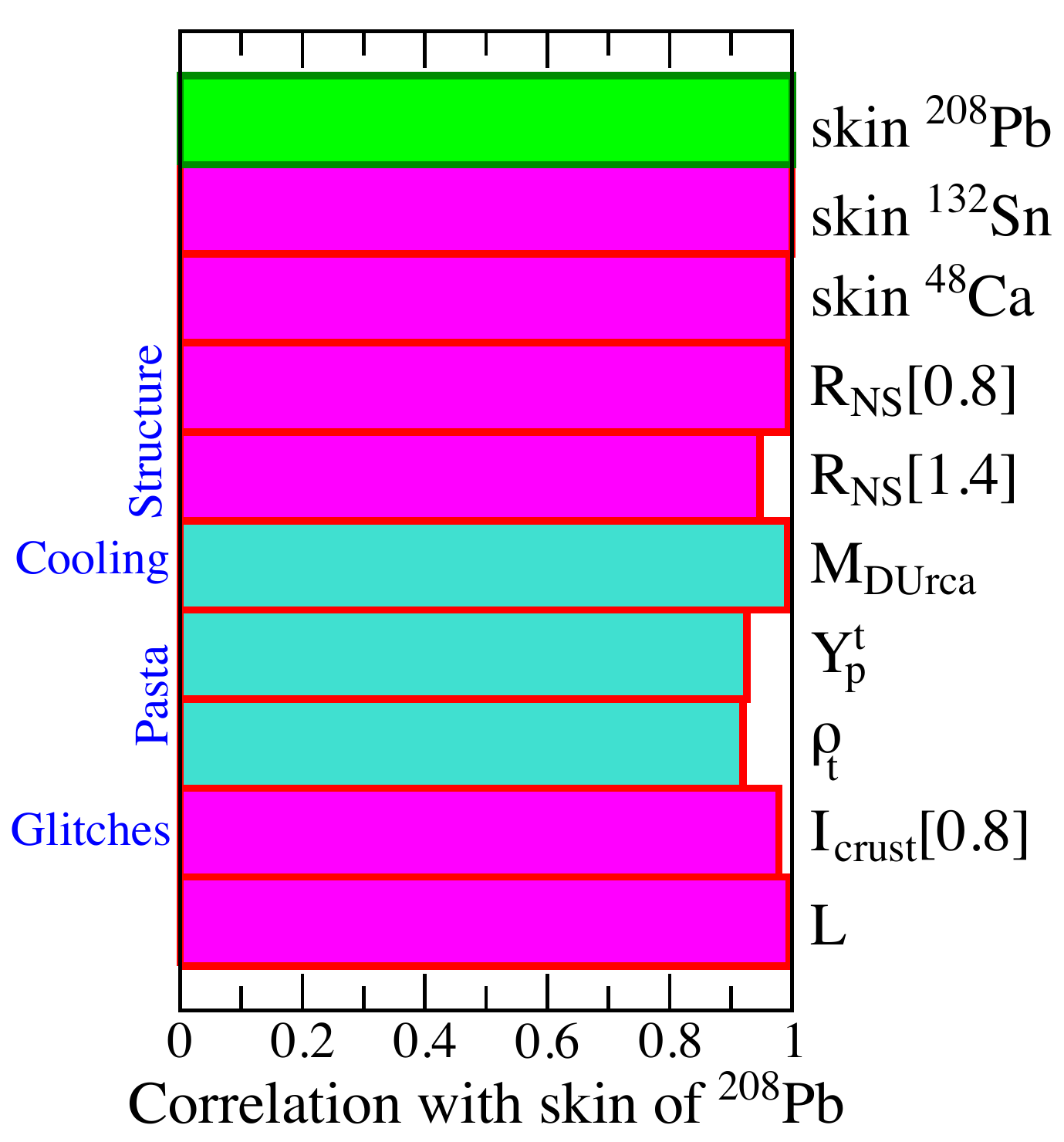}
 \vspace{-0.2cm}
  \caption{Correlation coefficients between $R_{\rm skin}^{208}$ and a variety of 
               neutron-star observables as predicted by the FSUGold density 
               functional\,\cite{Todd-Rutel:2005fa}.} 
 \label{Fig6}
 \end{center} 
\end{figure}

Recently, the Lead Radius Experiment (``PREX'') at the Jefferson Laboratory has 
provided the first model-independent evidence in favor of a neutron-rich skin in 
${}^{208}$Pb\,\cite{Abrahamyan:2012gp,Horowitz:2012tj}. Given that the $Z^{0}$
boson couples strongly to the neutron, parity violating electron scattering 
provides a clean probe of neutron densities that is free from strong-interaction 
uncertainties. As the proton radius of ${}^{208}$Pb is known extremely accurately 
from conventional (parity conserving) electron scattering, PREX effectively 
determined the neutron-skin thickness of  ${}^{208}$Pb to be: 
$R_{\rm skin}^{208}\!=\!{0.33}^{+0.16}_{-0.18}\,{\rm fm}$\,\cite{Abrahamyan:2012gp}. 
In a much anticipated and already approved follow-up experiment (``PREX-II'') the 
neutron-skin thickness of ${}^{208}$Pb will be measured with a significantly 
increased accuracy. The impact of such a measurement in constraining the 
value of $L$ is displayed with the green error bars in Fig.\,\ref{Fig5}. 

Perhaps surprisingly, the same pressure that is responsible for creating a neutron-rich 
skin in heavy nuclei is also responsible for supporting a neutron star against gravitational 
collapse. Thus, models that predict thicker neutron skins often produce neutron stars 
with larger radii\,\cite{Horowitz:2000xj,Horowitz:2001ya}. This makes possible to establish 
a powerful ``data-to-data'' relation between the neutron-rich skin of a heavy nucleus and 
the radius of a neutron star; a relation involving physical observables that differ by 18 
orders of magnitude! Such powerful connection is nicely illustrated in Fig.\,\ref{Fig6} 
which displays a strong correlation between $R_{\rm skin}^{208}$ and the stellar radius 
of a 0.8$M_{\odot}$ neutron star. In the case of a 1.4$M_{\odot}$ neutron star, the
correlation weakens slightly because, unlike low-mass neutron stars, medium-to-heavy 
neutron stars are also sensitive to the symmetry energies at densities above saturation 
density\,\cite{Carriere:2002bx}. Moreover, Fig.\,\ref{Fig6} also displays the enormous 
reach that an accurate measurement of $R_{\rm skin}^{208}$ could have on several
neutron-star properties\,\cite{Fattoyev:2012rm}; note that red bands imply a direct 
correlation whereas the blue bands an inverse correlation (i.e., an anticorrelation).
Indeed, an accurate measurement of $R_{\rm skin}^{208}$ could have a strong
impact on a variety of neutron-star observables related to their structure, cooling, 
and glitch mechanism\,\cite{Fattoyev:2010tb}. We underscore that all properties 
displayed in Fig.\,\ref{Fig6}---whereas pertaining to finite nuclei or neutron 
stars---were computed with a unique relativistic energy density functional. 

\section{Summary and Outlook}
\label{Summary} 

Neutron stars are truly gold mines for the study of physical phenomena
across a variety of disciplines ranging from elementary particle physics 
to general relativity. Indeed, binary pulsars have been used to conduct 
some of the most stringent tests of general relativity and their merger 
may provide both a powerful source of gravitational waves as well as a 
plausible site for r-process nucleosynthesis. Moreover, some of the 
fascinating phases predicted to exist in the neutron-star crust, such as
Coulomb crystals of neutron-rich nuclei and nuclear pasta, have counterparts
in condensed-matter physics. Finally, the stellar core may harbor new 
states of matter that may consist of deconfined quark matter, such as
color superconductors.  
 
Of course, from the perspective of nuclear physics, neutron stars hold
the answer to one of the most fundamental questions in the field:
\emph{How does subatomic matter organize itself and what phenomena 
emerge?} Indeed, this question figures prominently in the recent community 
report entitled ``Nuclear Physics: Exploring the Heart of 
Matter''. In this contribution we challenged the
original view of Baade and Zwicky that neutron stars consist of extremely 
closed packed neutrons\,\cite{Baade:1934}. Although undoubtedly such is
the most common perception of a neutron star, we showed that the 
reality is far different and much more interesting. In particular, during our
journey through a neutron star we uncovered a myriad of exotic states 
of matter that are speculated to exist in a neutron star. Moreover, we 
underscored the critical role that laboratory experiments will play in 
elucidating the fascinating nature of these exotic states. By the same
token, we highlighted the fundamental role that observation of neutron
stars using telescopes operating over an enormous range of frequencies
will have in constraining the equation of state of neutron-rich matter. As
such, neutron stars provide a powerful intellectual bridge between Nuclear 
Physics and Astrophysics. 

Whereas the main goal of the present contribution was an introduction to
the fascinating world of neutron stars, the powerful tool used for their 
study, namely, \emph{relativistic density functional theory} represents 
the overarching theme of this whole volume. Density functional theory
provides a powerful---and perhaps unique---framework for the accurate 
calculation of nuclear properties. Based on the seminal work by
Kohn and collaborators, DFT shifts the focus from the complicated 
many-body wave function to the much simpler one-body density.  By 
doing so, the formidable challenge of deducing the exact ground-state 
energy and one-body density from the many-body wave function 
``reduces'' to the minimization of a suitable functional of the density.
Whereas the implementation of DFT to the understanding of electronic
properties of materials (work for which Walter Kohn was recognized
in 1998 with the Nobel prize in chemistry) is firmly based on the 
Coulomb interaction, the situation is significantly more difficult in the 
nuclear physics domain. This is because the 
parameters of the nuclear density functional can not be computed 
directly from QCD. By necessity then, the parameters of the model 
must be directly calibrated (i.e., fitted) to physical observables. 
However, by directly fitting to data, the short-distance structure of 
the theory as well as many other complicated many-body effects 
get implicitly encoded in the parameters of the model. In this regard, 
DFT provides a unified and powerful approach that may be used to 
compute physical phenomena ranging over many distance scales,
such as the properties of finite nuclei and neutron stars; systems 
that differ in size by 18 orders of magnitude! Moreover, by implementing 
the calibration of the model parameters via a standard optimization 
procedure, DFT becomes a powerful microscopic theory that 
both predicts and provides well-quantified theoretical 
uncertainties\,\cite{PhysRevA.83.040001}. Documenting theoretical 
uncertainties is particularly critical as one extrapolates into uncharted 
regions of the observable landscape.

In this contribution we relied on relativistic DFT to explore three regions 
of the neutron star: the outer and inner crust as well as the outer core.
We determined that the outer stellar crust consists of a Coulomb crystal 
of neutron-rich nuclei embedded in a uniform electron gas. Remarkably,
the dynamics of the outer crust is only sensitive to the masses of both 
stable and unstable nuclei. Thus, mass measurements of exotic nuclei 
at rare isotope facilities will be instrumental in both constraining 
theoretical models as well as guiding the extrapolations into regions 
of the nuclear chart where mass measurement are unlikely to occur.
Moving into the inner crust, we established that at the top layers the
Coulomb crystal of neutron-rich nuclei is now in equilibrium with a 
dilute, and likely superfluid, neutron vapor. Hence, in this region one
is sensitive to the equation of state of dilute neutron matter, a topic 
of enormous interest and one that holds an intimate connection to
cold atomic Fermi gases in the unitary regime. Given that the EOS 
of neutron-star matter must span an enormous range of densities, a
study of dilute neutron matter can be used to further constrain the form 
of the relativistic density functional. As one moves even deeper into the 
inner crust, one discovers a complex pasta phase that displays unique 
and fascinating dynamical features. Although finding clear signatures of 
its existence has proved elusive, the lack of x-ray pulsars with long spin 
periods has recently been suggested as the first observable manifestation 
of the nuclear pasta phase. Recent molecular dynamics simulations seem 
to support such an assertion. 

Finally, we explored the deep stellar core which consists of a uniform quantum 
liquid of neutrons, protons, electrons and muons in chemical equilibrium. In this 
region the relativistic character of the density functional is particularly critical, 
as it provides a Lorentz covariant extrapolation to the high-density regime. That 
is,  the relativistic nature of the functional guarantees that the speed of sound in 
matter will always remain below the speed of light at all densities, a feature that 
is often not displayed by \emph{non-relativistic} density functionals. Whereas
one of the main goals of a relativistic density functional is to provide a unified
description of phenomena that happens at many distance scales, we particularly
highlighted the strong correlation between a fundamental property of finite 
nuclei---namely, the thickness of the neutron skin---and the radius of a neutron 
star. Thus, a laboratory 
measurement of the neutron-skin thickness of ${}^{208}$Pb (PREX) can constrain
the radius of a neutron star. Conversely, accurate measurements of stellar radii 
can help refine the functional. In particular, we discussed how a recent analysis 
of quiescent low-mass x-ray binaries seem to favor neutron stars with small
stellar radii. Such a finding, which seems to favor a soft EOS, poses
serious challenges to theoretical models that must simultaneously account for
massive neutron stars of at least 2 solar masses. Moreover, small stellar radii 
also appear at odds with the original PREX report of a fairly large neutron-skin 
thickness in ${}^{208}$Pb, albeit with large error bars. However, if future laboratory 
experiments and astronomical observations confirm that both $R_{\rm skin}^{208}$ 
is thick and stellar radii are small, this would strongly suggest a softening of the 
EOS due to the onset of a phase transition. Regardless,
the EOS must eventually stiffen again to account for the existence of massive 
neutron stars. Such extraordinary behavior will confirm the fundamental role of 
neutron 
stars as unique laboratories for the study of dense nucleonic matter. Independent
of the surprises that may lie ahead, relativistic density functional theory provides
the only tractable microscopic theory that can describe nuclear phenomena ranging 
from the physics of finite nuclei to the structure of neutron stars.
\bigskip\bigskip

\centerline{\bf Acknowledgments}
This material is based upon work supported by the United States Department 
of Energy Office of Science, Office of Nuclear Physics under Award Number 
DE-FD05-92ER40750.

\bibliography{WSBook.bbl}

\end{document}